# THE $^7$Li($n$, γ)$^8$Li RADIATIVE CAPTURE AT ASTROPHYSICAL ENERGIES

*S. B. Dubovichenko,*[*] *A. V. Dzhazairov-Kakhramanov*[†,§]

[§] Corresponding author E-mail: albert-j@yandex.ru

[*]*V. G. Fessenkov Astrophysical Institute "NCSRT" NSA RK, 050020, Observatory 23, Kamenskoe plato, Almaty, Kazakhstan; Institute of nuclear physics NNC RK, 050032, str. Ibragimova 1, Almaty, Kazakhstan, dubovichenko@mail.ru*
[†]*Institute of Nuclear physics NNC RK, 050032, str. Ibragimova 1, Almaty, Kazakhstan, albert-j@yandex.ru*

**Abstract.** The versions of intercluster interaction potentials describing the resonance nature of some phase shifts of the n$^7$Li elastic scattering at low energies and the $P_2$ ground state of $^8$Li in the n$^7$Li cluster channel have been constructed. The possibility of describing the total cross sections of $^7$Li(n, γ)$^8$Li within the energies from 5 meV (5·10$^{-3}$ eV) to 1 MeV, including resonance at 0.25 MeV, has been demonstrated for the potentials obtained in the potential cluster model with forbidden states.

**1 Introduction**

The $^7$Li(n, γ)$^8$Li radiative capture reaction at astrophysical energies with the formation of β-active $^8$Li is not a part of basic thermonuclear cycles [1-4], but it can play a certain role in some models of the Big Bang [5]. For example, it is believed that the primordial nucleosynthesis followed the basic reaction chain of the form [6]:

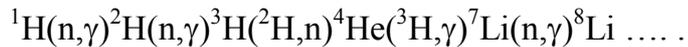

$^1$H(n,γ)$^2$H(n,γ)$^3$H($^2$H,n)$^4$He($^3$H,γ)$^7$Li(n,γ)$^8$Li …. .

Furthermore, the reaction under consideration is the mirror reaction to the $^7$Be(p, γ)$^8$B capture process, where $^8$B decays into $^8$Be+e$^+$+ν due to the weak interaction. Neutrinos from this reaction have rather high energy and have been registered on Earth over several decades, but the unstable $^8$Be nucleus decays into two α-particles. The $^7$Be(p, γ)$^8$B capture reaction is one of the final processes of the proton-proton chain [1], which apparently accounts for most of the energy release of the Sun and other stars of our Universe.

The potentials of n$^7$Li interaction in continuous and discrete spectra are required for calculation of the total cross sections of the radiative neutron capture on $^7$Li in the frame of the potential cluster model (PCM) [7-10] that takes into account forbidden states (FS). To find them, we assume that such potentials correspond to the classification of cluster states according to orbital symmetries [7], as we did it in earlier works [11-13] for other cluster nuclei. Therefore, in this article, we will start by considering this particular type of interaction for the n$^7$Li system.

The PCM potentials of scattering processes are usually constructed based on the elastic scattering phase shifts extracted from experimental differential cross sections using phase shift analysis. The bound state (BS) interactions are obtained by fulfilling the requirement to reproduce some characteristics of such states, considered in a certain cluster channel. It is assumed that the cluster channel comprising the initial particles in the reaction contributes the most to the BS [14]. $^8$Li is a stable nucleus with respect to strong interactions as it decays to $^8$Be due to weak forces only. Therefore, it is possible to consider this nucleus as the n$^7$Li cluster system and to use the known PCM methods [7,8] for its description.

## 2 Classification of the orbital cluster states

Note that n$^7$Li system has the isospin projection $T_z = -1$, which is only possible in the case of the total isospin $T = 1$ [15]. Therefore, this system, unlike the p$^7$Li system that is isospin-mixed with $T = 0$ and 1, is isospin-pure like the p$^7$Be system with $T_z = +1$ and $T = 1$ [16,17]. However, as in the p$^7$Li system, the spin can be equal to 1 and 2, and some states of the n$^7$Li system can also be mixed in spin.

It has been shown [8,16,18] that if scheme {7} is used for $^7$Li, then possible Young schemes {8} and {71} [19] for the system of eight particles turn out to be forbidden, because there cannot be more than four cells in a row [20-22]. Thus, they correspond to Pauli-forbidden states with relative motion moments $L = 0$ and 1, which are determined by the Elliot rule [20].

In the second case, when scheme {43} is chosen for $^7$Li, the n$^7$Li and p$^7$Be systems contain forbidden states in the $^3P$ waves with scheme {53} and in the $^3S_1$ wave with the WF symmetry {44}, and have the $P_2$ allowed state (AS) with the spatial scheme {431}, as is shown in Table 1 [8,16]. Therefore, the n$^7$Li potentials in the triplet spin state should have the forbidden bound $^3S_1$ state with scheme {44} for scattering processes, as well as the forbidden and allowed bound levels in the $^3P$ waves with the Young schemes {53} and {431}; with the latter corresponding to the $^3P_2$ ground bound state of $^8$Li in the n$^7$Li channel.

The allowed symmetries at $S = 2$ and consequently, the bounded allowed levels in the n$^7$Li system, are absent at any values of orbital moment $L$ [8,16,18,23]. Therefore, the potential of the $^5S_2$ scattering wave has the bound FS with scheme {44}, and in $^5P$ scattering waves, the potential has the FS with schemes {53} and {431}. The latter can be in the continuous spectrum and the potential has only one bound FS with scheme {53}. However, his conclusion is not the only possible outcome, and there can be a version of $^5P$ scattering potentials with two bound FS for schemes {53} and {431}.

It seems that as a third variant it is possible to consider both allowed Young schemes {7} and {43} for the ground state of $^7$Li, because both are among the FS and AS of this nucleus in the $^3$H$^4$He configuration [8,14]. Then the level classification will be slightly different, the number of FS will increase, and an additional forbidden bound level will appear in each partial wave with $L = 0$ and 1. This more complete scheme of orbital states is given in Table 1. It was considered in detail in earlier work [16] and in fact, represents the sum of the first and second cases described above [23]. The results for the first variant of the classification of FS, which is obtained by using only the orbital scheme {7} for $^7$Li, are shown in italics in the last four columns of Table 1.

**Table 1** Classification of the orbital states in N$^7$Li (N$^7$Be) systems.

| Systems | T | S | $\{f\}_T$ | $\{f\}_S$ | $\{f\}_{ST}=\{f\}_S\otimes\{f\}_T$ | $\{f\}_L$ | L | $\{f\}_{AS}$ | $\{f\}_{FS}$ |
|---|---|---|---|---|---|---|---|---|---|
| p$^7$Li n$^7$Be | 0 | 1 | {44} | {53} | {71} + {611} + {53} + {521} + {431} + {4211} + {332} + {3221} | *{8}* | 0 | - | *{8}* |
| | | | | | | *{71}* | 1 | - | *{71}* |
| | | | | | | {53} | 1,3 | - | {53} |
| | | | | | | {44} | 0,2,4 | - | {44} |
| | | | | | | {431} | 1,2,3 | {431} | - |
| | | 2 | {44} | {62} | {62} + {521} + {44} + + {431} + {422} + {3311} | *{8}* | 0 | - | *{8}* |
| | | | | | | *{71}* | 1 | - | *{71}* |
| | | | | | | {53} | 1,3 | - | {53} |
| | | | | | | {44} | 0,2,4 | - | {44} |
| | | | | | | {431} | 1,2,3 | - | {431} |
| p$^7$Be n$^7$Li p$^7$Li n$^7$Be | 1 | 1 | {53} | {53} | {8} + 2{62} + {71} + {611} + { 53} + {44} + + 2{521} + {5111} + {44} + {332} + 2{431} + 2{422} + {4211} + {3311} + {3221} | *{8}* | 0 | - | *{8}* |
| | | | | | | *{71}* | 1 | - | *{71}* |
| | | | | | | {53} | 1,3 | - | {53} |
| | | | | | | {44} | 0,2,4 | - | {44} |
| | | | | | | {431} | 1,2,3 | {431} | - |
| | | 2 | {53} | {62} | {71} + {62} + {611} + + 2{53} + 2{521} + 2{431} + {422} + {4211} + {332} | *{8}* | 0 | - | *{8}* |
| | | | | | | *{71}* | 1 | - | *{71}* |
| | | | | | | {53} | 1,3 | - | {53} |
| | | | | | | {44} | 0,2,4 | - | {44} |
| | | | | | | {431} | 1,2,3 | - | {431} |

*Note.* Here, $T$, $S$ and $L$ are isospin, spin and orbital momentums of two particles; $\{f\}_S$, $\{f\}_T$, $\{f\}_{ST}$ and $\{f\}_L$ are spin, isospin, spin-isospin [20] and possible orbital Young schemes; $\{f\}_{AS}$, $\{f\}_{FS}$ are Young schemes of the allowed and forbidden orbital states.

However, previous work [24] for the n$^6$Li system showed that it is possible to use the allowed scheme {42} for $^6$Li without taking into account its forbidden configuration {6}. Therefore, we will further consider the second variant of the FS structure and potentials with the allowed scheme {43} for $^7$Li as the main variant of the classification of FS and AS in such a system. Thus, we will consider that the potentials of the $^{3,5}S$ scattering waves, which are necessary for the consideration of the $E$1 electromagnetic transitions to the GS of $^8$Li at the neutron capture on $^7$Li, have the bound forbidden states with scheme {44}. The potential of the resonance $^5P_3$ scattering wave at 0.25 MeV, which allows considering $M$1 transition to the GS of $^8$Li, can have one bound FS with scheme {53} or two bound forbidden states with schemes {53} and {431}. The potential of the BS of $^8$Li in the n$^7$Li channel, which is the mixture of two $^3P_2$ and $^5P_2$ states, has one forbidden bound state with scheme {53} and one allowed bound state with scheme {431}, corresponding to the GS with the binding energy -2.03239 MeV [15].

## 3 Potential description of the elastic scattering

We have not managed to find any data on the elastic scattering phase shifts for the n$^7$Li or p$^7$Be systems at astrophysical energies [25]. Therefore, here we will construct the scattering potentials for the n$^7$Li system in analogy to the p$^7$Li scattering [16], based on the $^8$Li spectrum data [15], shown in Fig. 1, together with similar spectra of $^8$Be and $^8$B. The spectra are shown aligning the levels $2^+1$ of $^8$Li and $^8$B, which are stable in terms of nuclear interactions ground states, decaying only due to the weak forces.

The considered earlier bound state of the p$^7$Li system with $J^\pi T = 0^+0$ [16,26], corresponding to the ground state (GS) of $^8$Be, can only be formed in the triplet spin state of the p$^7$Li cluster system with $L = 1$, so it is a spin-pure $^3P_0$ state with $T = 0$ [15]. Accordingly, all previously obtained potentials of this system [16] corresponded to the triplet spin state with the above obtained number of AS and FS (see Table 1). All electromagnetic transitions take place between different levels in the triplet spin state, which according to Table 1, has the allowed Young scheme and consequently, the allowed bound state corresponding to the ground state of $^8$Be.

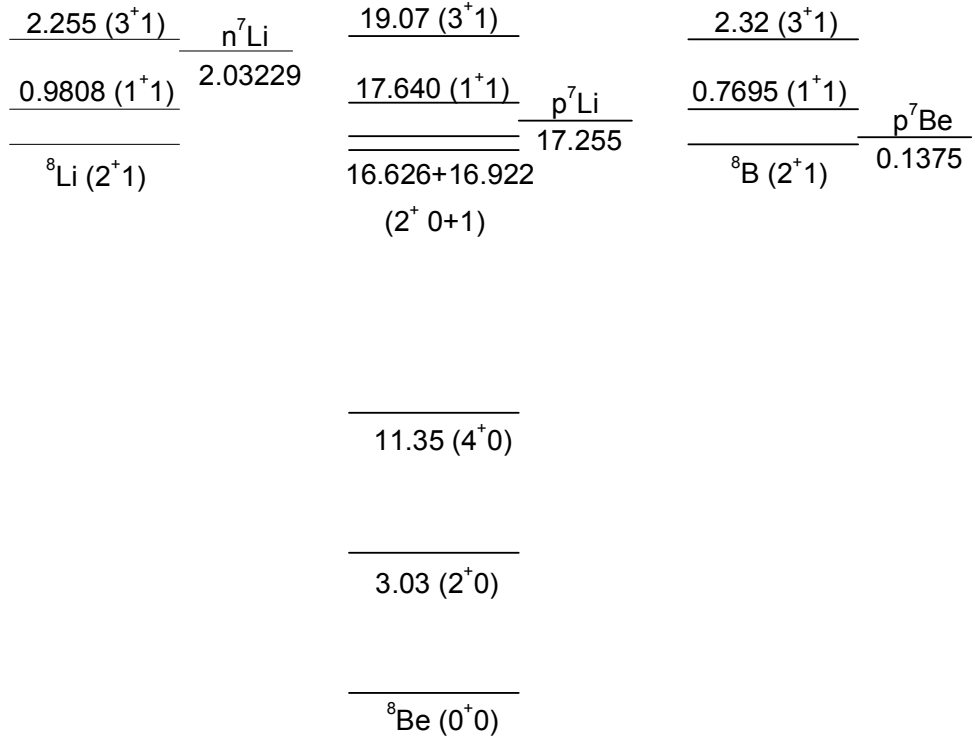

**Figure 1** Energy level spectrum in MeV (c.m.) of $^8$Li, $^8$Be and $^8$B [15].

In particular, we have considered the $E1$ transition between the $^3S_1$ scattering state with $T = 0$ and 1 and the $^3P_0$ ground state with $T = 0$, and the $M1$ process between the $^3P_1$ resonance wave (with $T = 1$) and the same $^3P_0$ state of $^8$Be, which take place with the isospin changes. Let us note that in the case of the p$^7$Li system with $T_z = 0$ some scattering states, which have no resonance with the experimentally measurable isospin, are isospin-mixed with $T = 0$ and 1. Because for the $^3S_1$ wave we considered the transitions with $\Delta T = 1$, only a part of the potential was obtained (with $T = 1$) [16]. For the $^3P_1$ wave, its potential describes the resonance level of $^8$Be with experimentally obtained isospin $T = 1$, and this state turns out to be isospin-pure. These two processes allow the complete description of the experimental data on the astrophysical *S*-factor of the radiative proton capture on $^7$Li with changing isospin, i.e., they fulfill the requirement $\Delta T = 1$ [16].

In this case, the bound state of the n$^7$Li system with $J^\pi T = 2^+1$, corresponding to the GS of $^8$Li, can be formed at $S = 1$ and 2 with the orbital moment $L = 1$, and is the mixture of the $^3P_2$ and $^5P_2$ states. However, there are no allowed Young schemes (allowed states) for $S = 2$. As this follows from the results of work [16] and from the above classification (see Table 1), we should take into account the presence in the GS of the $^5P_2$ wave admixture, which is necessary for consideration of the *M*1-transition from the $^5P_3$ resonance to the GS of $^8$Li.

The level $J^\pi T = 3^+1$ in the $^8$Li spectrum (see Fig. 1) corresponds to the resonance $^5P_3$ phase shift of the n$^7$Li elastic scattering at the energy of 0.22 MeV (c.m.) or 0.25 MeV (l.s.) above the n$^7$Li threshold [15]. Resonance $^5P_3$ state is spin-pure and can only be formed when the total spin $S = 2$, if we consider only the lowest possible orbital moments. Furthermore, we will use the data for energy levels of $^8$Li and the widths of those levels [15] for construction of the potential corresponding to the resonance of the n$^7$Li scattering phase shift.

The state with $J^\pi T = 1^+1$, characterized by $S = 1$, 2 and $L = 1$, is the $^{3+5}P_1$-level in the n$^7$Li channel, and is bound at the energy of 0.9808 MeV relative to the GS of $^8$Li, or at the energy of -1.05149 MeV relative to the n$^7$Li-channel threshold [15]. We will also consider the *E*1 transitions to this level from the triplet and quintet *S* scattering waves. Therefore, all further results will apply to the $^7$Li(*n*, $\gamma_0$)$^8$Li and $^7$Li(*n*, $\gamma_1$)$^8$Li reactions and the sum of their cross sections. Moreover, by analogy with the p$^7$Li scattering and based on data [15], we will consider that $^3S_1$ and $^5S_2$ phase shifts in the range up to 1 MeV are practically equal to zero. This is confirmed by the absence of resonance levels with negative parity in the spectrum of $^8$Li at such energies.

Because earlier in the p$^7$Li system [8] we considered the versions of the potentials with two FS, we will further use for comparison potentials in all partial scattering waves with different numbers of FS that are required for further calculations of the radiative capture. First, we will find the *S*- and *P*-potentials with two FS - the third variant of the classification of AS and FS (see Table 1), as it follows from the results given above, and then we will consider the versions with one (the second classification option) and zero FS, i.e., with their complete absence in each partial wave.

Because [7,8,20,21] Gaussian and Woods-Saxon potentials give almost identical results for 1*p*-shell nuclei, we will use a Gaussian potential of the form:

$$V(r) = -V_0 \cdot \exp(-\alpha r^2). \tag{1}$$

Here, $V_0$ and $\alpha$ are the parameters of the potential, which are determined based on the resonance characteristics of elastic scattering in certain partial waves, and characteristics of the bound state of the n$^7$Li system, i.e., the GS of $^8$Li.

Phase shifts vanishing to zero for the $^3S_1$ and $^5S_2$ waves at low energies are obtained with the abovementioned potential (1) and parameters

$$V_S = 145.5 \text{ MeV and } \alpha_S = 0.15 \text{ fm}^{-2}. \tag{2}$$

Here, we will consider this version of the potential, because a similar potential was used in considering the p$^7$Li scattering in the $^3S_1$ state [16] - it contains two bound FS (as follows from the classification of states given in Table 1 for the third version) with the orbital schemes {8} and {44}.

The zero phase shift can also be obtained with the potential

$$V_S = 50.5 \text{ MeV and } \alpha_S = 0.15 \text{ fm}^{-2}, \tag{3}$$

which has only one bound FS for the second version of classification with scheme {44}, as well as with zero depth of the potential of Eq. (1) without FS, i.e., with $V_0 = 0$ for both $S$ waves of scattering.

Certainly, the near-zero $S$ phase shifts can be obtained with the help of other variants of parameters of the potential with one or two FS. In this case, it is impossible to fix parameters of such a potential unambiguously, and other combinations of $V_0$ and $\alpha$ are possible for Eqs (2) and (3).

The resonance $^5P_3$ phase shift of the n$^7$Li elastic scattering can be described with a Gaussian potential from Eq. (1) with parameters

$$V_P = 4967.45 \text{ MeV and } \alpha_P = 3.0 \text{ fm}^{-2}. \tag{4}$$

This potential has two bound forbidden states, which can be compared with schemes {53} and {431} for the second variant of the FS classification from Table 1, if we consider that the second FS with scheme {431} is the bound state. The calculation results for the $^5P_3$ scattering phase shift are shown in Fig. 2 with a dotted line. The resonance is at the energy of 254 keV (l.s.) with the width of 37 keV (c.m.), which coincides perfectly with the experimental value of 254(3) keV [15].

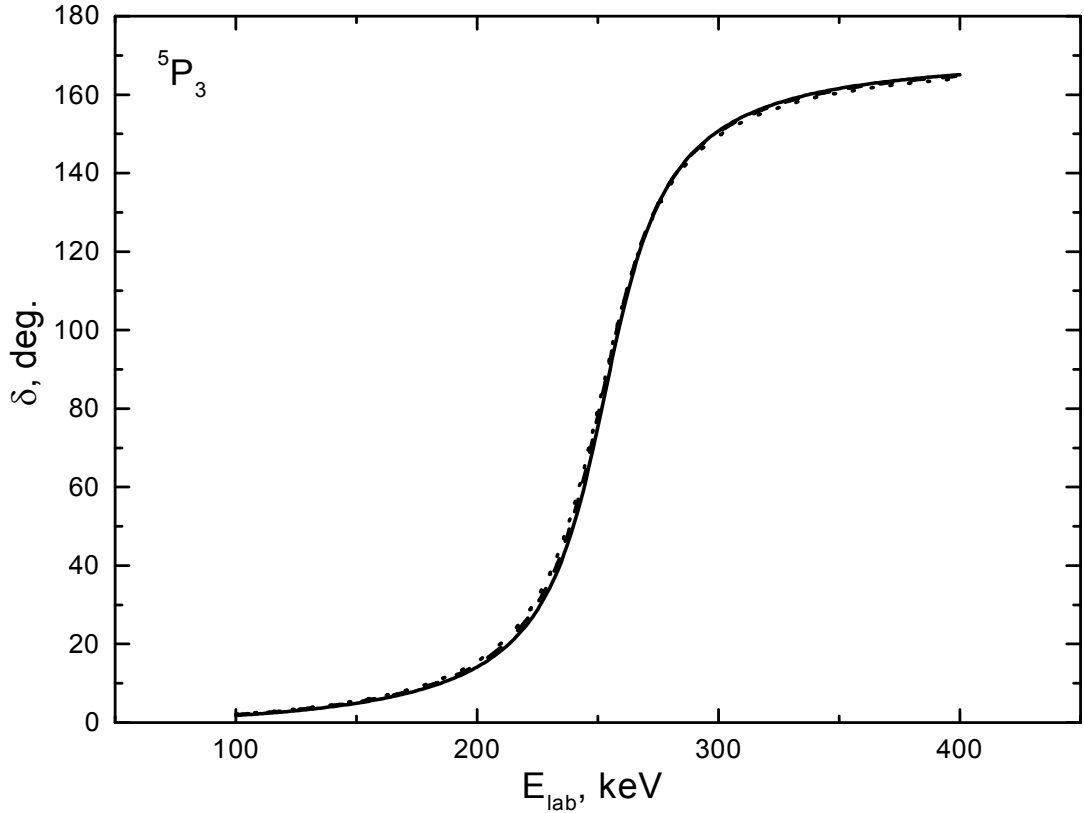

**Figure 2** The resonance $^5P_3$ phase shift of the n$^7$Li elastic scattering at low energies. Lines: Dotted line – phase shift for the potential of Eq. (4). Dashed line – for the potential of Eq. (5). Solid line – for the potential of Eq. (6). *Note*: the dashed line is almost hidden by the solid line.

The parameters of the potential with one forbidden state with {53}, which also correspond to the second variant of the classification, provided that the FS with scheme {431} is in the continuous spectrum, have the form

$$V_P = 2059.75 \text{ MeV and } \alpha_P = 2.5 \text{ fm}^{-2}. \qquad (5)$$

The results of the phase shift calculation are shown in Fig. 2 by the dashed line – the resonance is obtained at the energy of 254 keV. The width of the $^5P_3$ resonance is equal to 35 keV (c.m.), with the experimental values being equal to 35(5) or 33(6) keV (c.m.) according to different data from review [15].

The parameters of the potential without FS, which will be considered additionally, are represented as

$$V_P = 425.1 \text{ MeV and } \alpha_P = 1.5 \text{ fm}^{-2}. \qquad (6)$$

The calculation results of the $^5P_3$ phase shift with these parameters are shown in Fig. 2 by the solid line – the resonance is obtained at the energy of 255 keV and its width is equal to 34 keV (c.m.). It should be stressed here, that the parameters of this potential at the given number of the bound FS can be determined unambiguously according to the resonance energy and its width.

Because we will consider the second variant of the cluster classification for the GS of $^8$Li, the following parameters of the potential for the bound $^{3+5}P_2$ state of the n$^7$Li system, which corresponds to the GS of $^8$Li in the cluster channel under consideration, can be used:

$$V_{GS} = 429.383779 \text{ MeV and } \alpha_{GS} = 0.5 \text{ fm}^{-2}. \qquad (7)$$

In addition to the allowed bound state corresponding to the GS of $^8$Li with {431}, such $^{3+5}P_2$ potential has the bound FS with {53}. The binding energy of -2.0322900 MeV, which coincides completely with the experimental value [15], the charge radius of 2.38 fm and the mass radius of 2.45 fm have been obtained with this potential based on the finite-difference method (FDM) [11,26] with an accuracy of $10^{-7}$ MeV. Apparently, the root-mean-square charge radius of $^8$Li should not be substantially more than that of $^7$Li, which is equal to 2.35(10) fm [27]. Therefore, the value of the root-mean-square charge radius obtained above, in the n$^7$Li channel for the GS of $^8$Li, has a reasonable value. The zero value was used for the neutron charge radius, and its mass radius was taken to be equal to the corresponding radius of the proton 0.8775(51) fm.

The asymptotic constant (AC), which characterizes the behavior of the wave function of a bound state at large distances, for this potential of the GS turned out to be equal to $C_W = 0.78(1)$. The AC error is determined by its averaging in the range from 4 to 20 fm, where its value remains relatively stable. The asymptotic constant $C_W$ itself is defined as [28,29]

$$\chi_L(R) = \sqrt{2k_0} C_W W_{-\eta_L, L+1/2}(2k_0 R),$$

where: $\chi_L$ is the numerical wave function of the bound state obtained from the solution of the radial Schrödinger equation and normalized to unity; $W_{-\eta,L+1/2}(2k_0 R)$ is the Whittaker function of the bound state determining the asymptotic behavior of the wave function (WF), which is the solution of the same equation without nuclear potential, i.e., long distance solution; $k_0$ is the wave number determined by the channel binding energy; η is the Coulomb parameter, which in this case equals zero; $L$ is the orbital moment of the bound state.

Note, that for neutral particles [30]

$$W_{0,L+1/2}(z) = e^{-z/2} \sum_{n=0}^{L} z^{-n} \frac{(L+n)!}{(L-n)!n!} .$$

For comparison, we would like to give here the asymptotic constant value obtained in work [31], which equals $C(p_{3/2}) = 0.62$ fm$^{-1/2}$, or 0.81 after recalculation to the dimensionless value with $\sqrt{2k_0} = 0.767$. In [32], the values obtained were 0.59 fm$^{-1/2}$ for $C(^5P_2)$ and 0.28 fm$^{-1/2}$ for $C(^3P_2)$, which in dimensionless form gives 0.77 and 0.36. Note that a slightly different definition of AC was used in [31,32]

$$\chi_L(R) = C_W W_{-\eta,L+1/2}(2k_0 R) .$$

The following parameters for the first excited state (FES) were obtained:

$V_{ES} = 422.126824$ MeV and $\alpha_{ES} = 0.5$ fm$^{-2}$. (8)

The allowed BS with scheme {431} here corresponds to the first excited state of $^8$Li. In addition, this $^{3+5}P_1$ potential has the FS in full accordance with the second version of the classification of orbital states, as shown in Table 1. The binding energy of -1.051490 MeV, fully coinciding with the experimental value [15], the charge radius of 2.39 fm and the mass radius of 2.52 fm have been obtained with this potential based on the FDM [11,26] with an accuracy of $10^{-6}$ MeV. The asymptotic constant for this potential was equal to $C_W = 0.59(1)$. The asymptotic constant error was determined by its averaging in the range of 4–25 fm, where AC remains relatively stable.

For additional control of the bound energy calculations, the variational method (VM) with the expansion of the cluster wave function of the relative motion of the n$^7$Li system on a non-orthogonal Gaussian basis was used [8,33]

$$\Phi_L(R) = \frac{\chi_L(R)}{R} = R^L \sum_i C_i \exp(-\beta_i r^2) ,$$

where $\beta_i$ – variational parameters and $C_i$ – WF expansion coefficients.

With the dimension of the basis $N = 10$, the energy of -2.0322896 MeV was obtained for the GS potential (7). The residuals have the order of $10^{-14}$ [33], the asymptotic constant at the range 5–15 fm equals 0.78(1), and the charge radius does not differ from the previous FDM results. The expansion parameters of the received variational GS radial wave function of $^8$Li in the n$^7$Li cluster channel are shown in

Table 2.

**Table 2** The coefficients and expansion parameters of the radial variational wave function of the ground state of $^8$Li for the n$^7$Li channel in non-orthogonal Gaussian basis [8].

| $i$ | $\beta_i$ | $C_i$ |
|---|---|---|
| 1 | 2.111922863906128E-001 | -1.327201117117602E-001 |
| 2 | 1.054889049037163E-001 | -4.625421860118692E-002 |
| 3 | 9.251179926861837E-003 | -1.875176301729967E-004 |
| 4 | 2.236449875501786E-002 | -2.434284188136483E-003 |
| 5 | 4.990617934603718E-002 | -1.282820835431680E-002 |
| 6 | 3.849142988488459E-001 | -2.613687472261875E-001 |
| 7 | 5.453825421384008E-001 | -2.108830320871615E-001 |
| 8 | 1.163891769476509 | 1.438162032150163 |
| 9 | 1.716851806191120 | 1.426517649534997 |
| 10 | 2.495389760080367 | 1.792643814712334E-001 |

*Note.* The normalization factor of the wave function is $N$ = 9.999998392172028E-001.

Because the variational energy decreases as the dimension of the basis increases, yielding the upper limit of the true binding energy, and because the finite-difference energy increases as the step size decreases and the number of steps increases [33], the average value of -2.0322898(2) MeV can be taken as a realistic estimate of the binding energy in this potential. Therefore, the real accuracy of determination of the binding energy of $^8$Li in the n$^7$Li cluster channel for this potential, using two methods and two different computer programs, is at the level ±0.2 eV.

The energy of -1.051488 MeV was obtained using the variational method for the first excited state with residuals of $10^{-14}$ – none of the other characteristics differs from the ones obtained above based on the finite-difference method. The expansion parameters of the wave function are shown in Table 3, and the average energy can be written as -1.051489(1) MeV, i.e., the calculation error for such a potential is equal to 1 eV and agrees with the fixed FDM accuracy of $10^{-6}$ MeV.

**Table 3** The coefficients and expansion parameters of the radial variational wave function of the first exited state of $^8$Li for the n$^7$Li channel in non-orthogonal Gaussian basis [8].

| $i$ | $\beta_i$ | $C_i$ |
|---|---|---|
| 1 | 2.034869839899546E-001 | -1.268995424220545E-001 |
| 2 | 9.605255016688968E-002 | -4.250984818616291E-002 |
| 3 | 6.473027608029138E-003 | -2.029700124120304E-004 |
| 4 | 1.743880699865412E-002 | -2.308434897721290E-003 |
| 5 | 4.241481028548091E-002 | -1.167539819061673E-002 |
| 6 | 3.943411589808715E-001 | -2.876208138367455E-001 |
| 7 | 5.758070107927670E-001 | -1.307197681388061E-001 |
| 8 | 1.148526246366072 | 1.335023264621784 |
| 9 | 1.706295940575450 | 1.303208908841006 |
| 10 | 2.491484117851039 | 1.558051077479201E-001 |

*Note.* The normalization factor of the wave function is $N$ = 9.999907842436313E-001.

# 4 Cross section of the $^7$Li$(n, \gamma)^8$Li capture reaction

First, we would like to note that apparently the $E1$ transition in this system was first considered in works [34,35], where the possibility of the correct description of the total cross sections in non-resonance energy range was shown based on one particle model with the Woods-Saxon potential matched with energy levels of $^8$Li. Furthermore, this process was considered based on a direct capture model, for example, in [36]. Similar results were achieved in more recent works [37,38], where a correct description of the total capture cross sections based on the $E1$ process for non-resonance energies. As far as we know, based on model independent methods, results with an acceptable description of the resonance at 0.25 MeV were only obtained recently [39].

Furthermore, we will show that similar results in describing this resonance based on the $M1$ transition from the $^5P_3$ scattering wave, which has a resonance at this energy, to the $^5P_2$ component of the GS WF of $^8$Li, can be obtained in the potential cluster model [16,26,40,41]. As we did earlier for the proton capture on $^7$Li [16], when considering the electromagnetic transitions in the $^7$Li$(n,\gamma_0)^8$Li reaction, we will take into account the $E1$ transition from the $^3S_1$ scattering wave to the triplet $^3P_2$ part of the GS WF. In addition, compared with the p$^7$Li system [16], there will be an additional transition from the quintet $^5S_2$ scattering wave to the quintet $^5P_2$ part of wave function of the GS of $^8$Li. The $E1$ process from both $^{3+5}S$ scattering waves to the first excited $^{3+5}P_1$ state should be taken into account.

Thus, the total cross section of the whole process is represented as

$$\sigma_0(E1+M1) = \sigma(E1, {}^3S_1 \to {}^3P_2) + \sigma(E1, {}^5S_2 \to {}^5P_2) + \sigma(M1, {}^5P_3 \to {}^5P_2)$$

and

$$\sigma_1(E1) = \sigma(E1, {}^3S_1 \to {}^3P_1) + \sigma(E1, {}^5S_2 \to {}^5P_1)$$

The calculation results were compared with the experimental measurements of the total cross sections of the radiative capture reaction in the energy range from 5 meV to 1.0 MeV, laid down in works [5,42-47].

The expressions for the total radiative capture cross sections $\sigma(NJ,J_f)$ for $EJ$ and $MJ$ transitions in the potential cluster model are given, for example, in works [23,48] and are written as

$$\sigma_c(NJ, J_f) = \frac{8\pi K e^2}{\hbar^2 q^3} \frac{\mu}{(2S_1+1)(2S_2+1)} \frac{J+1}{J[(2J+1)!!]^2} A_J^2(NJ,K) \sum_{L_i,J_i} P_J^2(NJ,J_f,J_i) I_J^2(J_f,J_i)$$

where: $\sigma$ is the total cross section of the radiative capture process, $\mu$ is the reduced mass of the particles in the initial channel; $q$ is the wave number of the initial channel particles; $S_1$, $S_2$ are spins of the initial channel particles; $K$, $J$ are wave number and angular moment of $\gamma$-quantum in the final channel; $N$ represents $E$ or $M$ transitions of $J$-th multipolarity from the initial ($i$) to the final $J_f$ state, and for the electric orbital $EJ(L)$ transitions ($S_i = S_f = S$) we have

$$P_J^2(EJ, J_f, J_i) = \delta_{S_iS_f}[(2J+1)(2L_i+1)(2J_i+1)(2J_f+1)](L_i 0 J 0 | L_f 0)^2 \begin{Bmatrix} L_i & S & J_i \\ J_f & J & L_f \end{Bmatrix}^2,$$

$$A_J(EJ, K) = K^J \mu^J \left( \frac{Z_1}{m_1^J} + (-1)^J \frac{Z_2}{m_2^J} \right), \qquad I_J(J_f, J_i) = \langle \chi_f | R^J | \chi_i \rangle.$$

Here: $L_f$, $L_i$, $J_f$, $J_i$ are the angular moments of particles in the initial (*i*) and final (*f*) channels; $m_1$, $m_2$, $Z_1$, $Z_2$ are the masses and charges of the particles in the initial channel, respectively; $I_J$ is the integral over wave functions of the initial $\chi_i$ and final $\chi_f$ states as functions of relative cluster motion with the intercluster distance $R$. Let us stress that in our calculations, here and previously, we never used such a notion as a spectroscopic factor (see, e.g., [48]), i.e., its value was simply taken to be equal to unity.

For the spin part of the magnetic process $M1(S)$ in the model used, the following expression ($S_i = S_f = S$, $L_i = L_f = L$) [16] was obtained

$$P_1^2(M1, J_f, J_i) = \delta_{S_iS_f} \delta_{L_iL_f}[S(S+1)(2S+1)(2J_i+1)(2J_f+1)] \begin{Bmatrix} S & L & J_i \\ J_f & 1 & S \end{Bmatrix}^2,$$

$$A_1(M1, K) = \frac{e\hbar}{m_0 c} K \sqrt{3} \left[ \mu_1 \frac{m_2}{m} - \mu_2 \frac{m_1}{m} \right],$$

$$I_J(J_f, J_i) = \langle \chi_f | R^{J-1} | \chi_i \rangle, \qquad J=1.$$

Here, $m$ is the mass of the nucleus, $\mu_1$ and $\mu_2$ are the magnetic moments of clusters, and the other symbols are the same as in the previous expression. The following values were used for magnetic moments of the neutron and $^7$Li: $\mu_n = -1.91304272\mu_0$ and $\mu(^7\text{Li}) = 3.256427\mu_0$ [49], with the absolute value of the magnetic moment being used for the neutron. The correctness of the expression for the $M1$ transition was pre-tested on the basis of the results obtained in our previous studies [12,16,40] for the proton radiative capture reactions on $^2$H and $^7$Li at low energies. Exact values of particle masses [50] were used in all calculations, and the constant $\hbar^2/m_0$ was taken to be equal to 41.4686 MeV fm$^2$.

Because three variants of the potentials for each partial wave were given, we will not dwell on the results for each of these combinations. Instead, we will immediately give the final and apparently, the best result for the calculation of the summarized by all transitions total cross section for the radiative neutron capture on $^7$Li to the GS at the energy up to 1 MeV (l.s.), which is shown by the dot-dashed line in Fig. 3a. The results for total cross sections, which take into account all processes at the GS and FES, are shown by the solid line.

These results of the total cross section calculations were obtained for the potential of the GS (7), for the $S$ scattering waves with parameters from (3), and for the potential of the $^5P_3$ resonance with parameters (5). The dashed line shows the cross section corresponding to the sum of the $E1$ transitions from the $^3S_1$ and $^5S_2$ waves to the GS, the dotted line shows the cross section of the $M1$ transition between the $^5P_3$ scattering state and the GS of $^8$Li. The potential (8) and the same potentials of the $^{3+5}S$ scattering waves (3) were used for the FES.

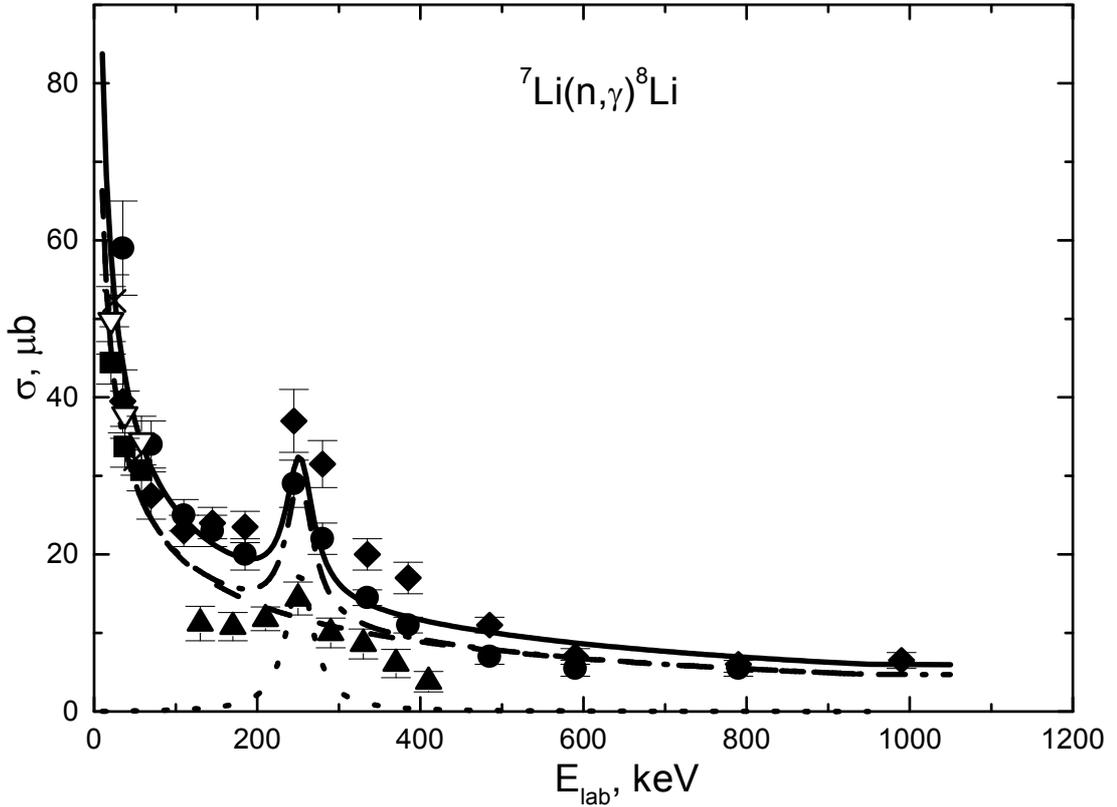

**Fig. 3a.** The total cross sections for the radiative neutron capture on $^7$Li. The experimental data: • and ♦ are from [42], ■ – [43] for capture to the GS, and ∇ - summarized cross sections for capture to the GS and FES, ▲ – [44], x – [45]. Dashed line – the $E$1 transitions from the $^3S_1$ and $^5S_2$ waves to the GS. Dotted line – the $M$1 transition between the $^5P_3$ scattering state and the GS of $^8$Li. Dot-dashed line – the calculation of the summarized by all transitions total cross section to the GS. Solid line – the results for total cross sections, which take into account all processes at the GS and FES.

A more exact shape and the values of the calculated total cross sections for these variants of the potentials at energies from 1 meV (1·10$^{-3}$ eV) to 150 keV, are shown in Fig. 3b. The results for total cross sections in accordance with all processes at GS and FES are shown by the solid line, whilst for transitions only to the GS of $^8$Li – by the dot-dashed line. It is clear from these figures that by using such potentials it is possible to describe the experimental data in the widest possible range of energies - from 5 meV to 1.0 MeV. It should be noted that the results, shown in Fig. 3b by the open circles, are the cross sections that were measured only for the capture to the GS of $^8$Li in work [46]. The measurements for capture to the GS (black squares) as well as the total cross sections taking into account transitions to the GS and FES (reverse open triangles) were made in the work [43]. Fig. 3c illustrates in detail the calculation results and experimental data for the energy range 20–60 keV. The crosses here show the data from work [45] for transitions $\gamma_0+\gamma_1$, while the curves are as in the previous figures.

The use of the variant of scattering potential for the $^5P_3$ wave with two FS (4) practically does not change the results of the description of the resonance at 0.25 keV. Therefore, the ambiguity of the FS number noted above does not affect the results. The calculation of the cross sections with the interaction without FS (6) leads to a notable decrease of the cross section value at the resonance energy, i.e., to a deterioration of the quality of the cross section description in this energy range.

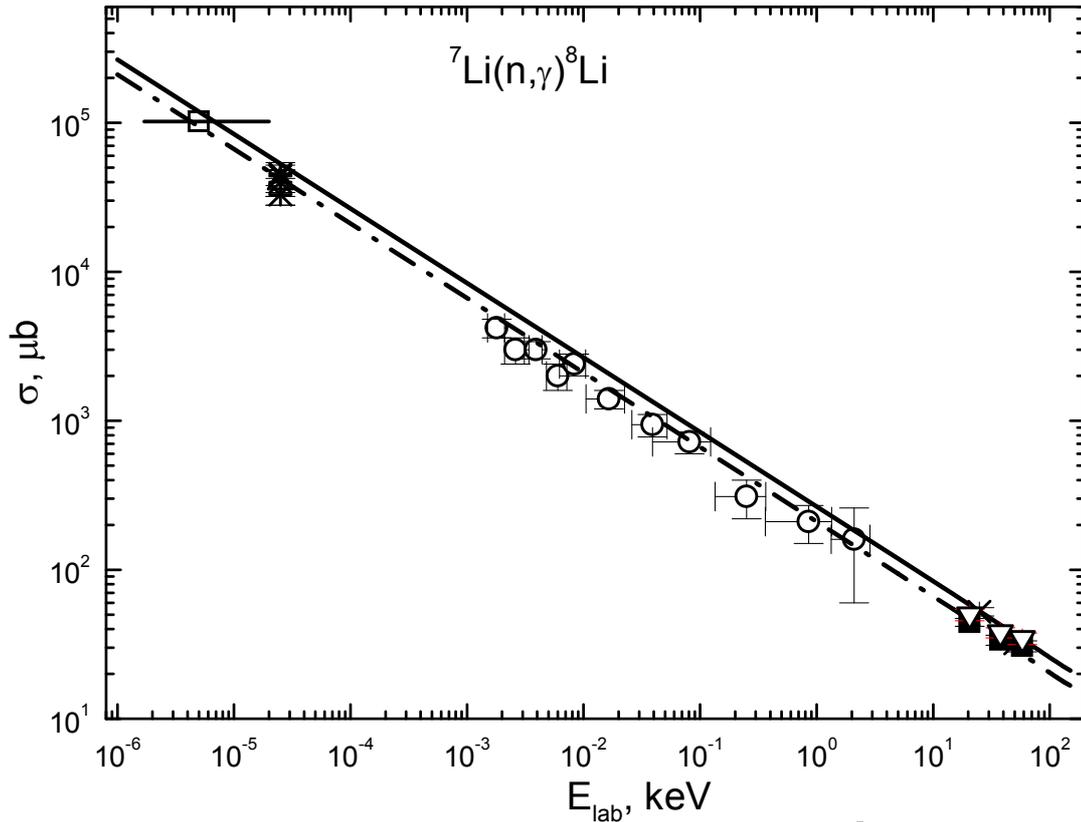

**Fig. 3b.** The total cross sections for the radiative neutron capture on $^7$Li at ultralow energies. Experiment: ■ – [43] (to the GS) and ∇ - summarized cross sections for capture to the GS and FES, ○ – [46], □ – [5], the solid horizontal line - the energy range where the measurements were made [5], Δ – [42], * – [47] and other data given in it, x – [45]. Lines: as in Fig. 3a.

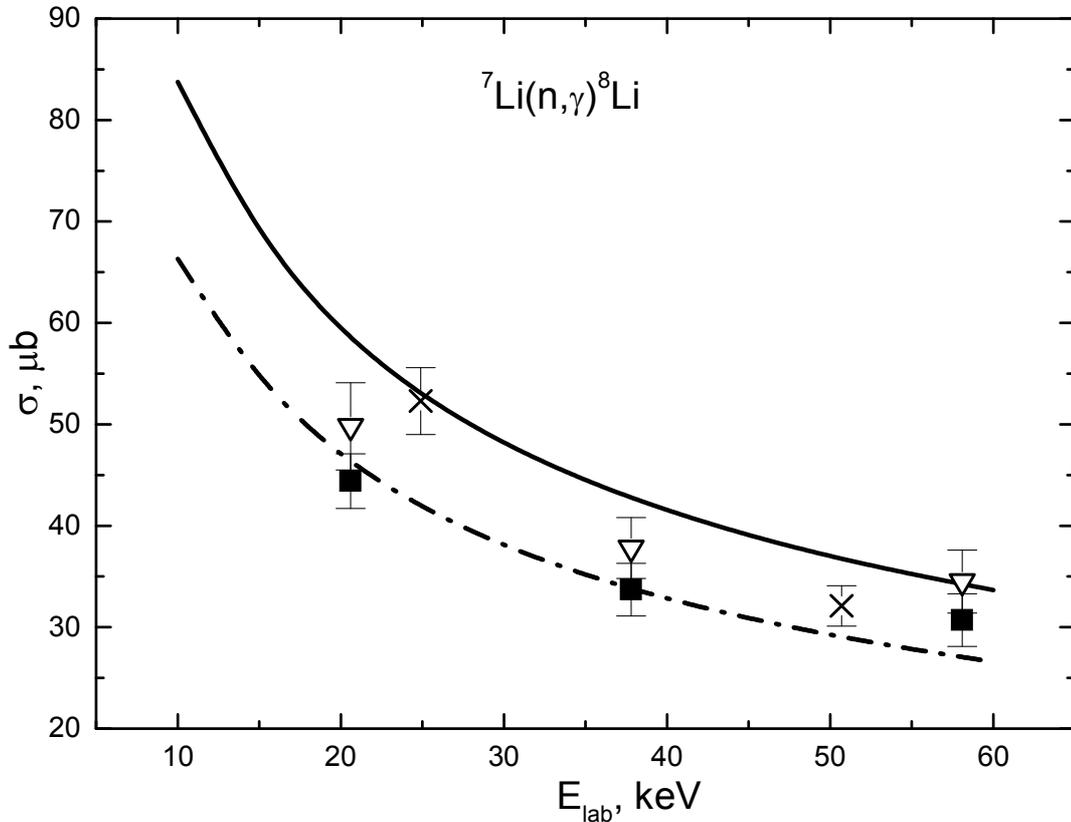

**Fig. 3c.** The total cross sections for the radiative neutron capture on $^7$Li. Experimental data: ■ – [43] for capture to the GS and ∇ – summarized cross sections for capture to the GS and FES, x – [45] for total capture cross sections to the GS and FES. Lines: as in Fig. 3a.

The variants of scattering potentials in the *S* waves with two FS or without FS produce almost no difference in the cross section calculation results. Even doubling the width of *S* potentials, notably to 0.3 fm$^{-2}$, with the depth of 100 MeV, has only a very slight effect on the value of the calculated cross sections. Only near-zero values of the scattering phase shifts at spin *S* = 1 and 2 are significant for these partial waves.

Consequently, we can assume that the use of the GS and FES potentials of $^8$Li in the n$^7$Li channel with one FS (7), (8) and the corresponding scattering potentials from Eqs (3) and (5), leads to a reasonable description of the available experimental data of total cross sections of radiative capture process for all considered energy regions, for the span between the upper and the lower limits, which is of the order of 10$^9$. The presence or absence of the FS in the potentials of *S* scattering waves does not play any role; only zero (0°±2°) values of the scattering phase shifts are important. Small changes in the value of the calculated total cross sections of the capture reaction do not allow definite conclusions to be drawn with respect to the number of the FS for the $^5P_3$ scattering potential in the resonance energy range.

Because at energies up to 100 keV the calculated cross section is almost a straight line (see Fig. 3b, solid line), it can be approximated by a simple function of the form:

$$\sigma(\mu b) = \frac{265.7381}{\sqrt{E_n(\text{keV})}}.$$

The value of constant 265.7381 μb·keV$^{1/2}$ was determined by a single point at a cross section with minimal energy of 1·10$^{-6}$ keV. The absolute value of the relative deviation: $M(E) = \left|[\sigma_{ap}(E) - \sigma_{theor}(E)]/\sigma_{theor}(E)\right|$, of the calculated theoretical cross sections, and the approximation of this cross section by the expression given above at energies less than 100 keV, does not exceed 1.5%.

Apparently, it can be suggested that this form of the total cross section energy dependence will be retained at lower energies. Therefore, we can perform an evaluation of the cross section value; for example, at the energy of 1 μeV (10$^{-6}$ eV=10$^{-9}$ keV) – it gives result 8.4 b.

The cross section approximation coefficient for the dot-dashed curve in Fig. 3b is equal to 210.538 μb·keV$^{1/2}$, and the cross section value at 1 μeV equals 6.7 b.

## 5 Conclusion

The *E*1 and *M*1 transitions from the $^{3+5}S_{1+2}$ and $^5P_3$ scattering waves to the $^{3+5}P_2$ bound state of $^8$Li in the n$^7$Li channel are considered in the potential cluster model [26,41,51]. It is quite possible to describe correctly the value of the total cross section in the non-resonant energy region, and even the position and value of the resonance in the neutron capture reaction on $^7$Li.

Note, that there are already eighteen cluster systems, the characteristics of which are given in Table 4 (properties of nuclei are from reviews [15,52-55]), considered by us earlier based on the potential cluster model with the classification of the orbital states according to the Young schemes [9,12,14,16,17,26,40,51,56-72], in terms of which it is possible to obtain acceptable results of the description of the radiative capture process characteristics.

**Table 4.** Characteristics of nuclei and cluster systems (spin, isospin, and parity), and references to works in which the astrophysical *S*-factors for radiative capture process were considered based on the potential cluster model.

| No. | Nucleus ($J^\pi$, $T$) | Cluster channel | $T_z$ | $T$ | Ref. |
|---|---|---|---|---|---|
| 1. | $^3$H ($1/2^+$,1/2) | n$^2$H | -1/2 + 0 = -1/2 | 1/2 | [62,66] |
| 2. | $^3$He ($1/2^+$,1/2) | p$^2$H | +1/2 + 0 = +1/2 | 1/2 | [12,26,40,51,67] |
| 3. | $^4$He ($0^+$,0) | p$^3$H | +1/2 - 1/2 = 0 | 0 + 1 | [17,26,41,51] |
| 4. | $^6$Li ($1^+$,0) | $^2$H$^4$He | 0 + 0 = 0 | 0 | [14,26,41] |
| 5. | $^7$Li ($3/2^-$,1/2) | $^3$H$^4$He | -1/2 + 0 = -1/2 | 1/2 | [14,26,41] |
| 6. | $^7$Be ($3/2^-$,1/2) | $^3$He$^4$He | +1/2 + 0 = +1/2 | 1/2 | [14,26,41] |
| 7. | $^7$Be ($3/2^-$,1/2) | p$^6$Li | +1/2 + 0 = +1/2 | 1/2 | [26,56,57,67] |
| 8. | $^7$Li ($3/2^-$,1/2) | n$^6$Li | -1/2 + 0 = -1/2 | 1/2 | [63,66] |
| 9. | $^8$Be ($0^+$,0) | p$^7$Li | +1/2 - 1/2 = 0 | 0 + 1 | [16,17,26,51,67] |
| 10. | $^8$Li ($2^+$,1) | n$^7$Li | -1/2 - 1/2 = -1 | 1 | Present paper |
| 11. | $^{10}$B ($3^+$,0) | p$^9$Be | +1/2 - 1/2 = 0 | 0 + 1 | [26,51,58] |
| 12. | $^{10}$Be ($0^+$,1) | n$^9$Be | -1/2 - 1/2 = -1 | 1 | [72] |
| 13. | $^{13}$N ($1/2^-$,1/2) | p$^{12}$C | +1/2 + 0 = +1/2 | 1/2 | [9,26,51,67] |
| 14. | $^{13}$C ($1/2^-$,1/2) | n$^{12}$C | -1/2 + 0 = -1/2 | 1/2 | [64,66] |
| 15. | $^{14}$N ($1^+$,0) | p$^{13}$C | +1/2 - 1/2 = 0 | 0 + 1 | [59,60,67,70] |
| 16. | $^{14}$C ($0^+$,1) | n$^{13}$C | -1/2 - 1/2 = -1 | 1 | [64,66] |
| 17. | $^{15}$C ($1/2^+$,3/2) | n$^{14}$C | -1/2 – 1 = -3/2 | 3/2 | [68] |
| 18. | $^{15}$N ($1/2^-$,1/2) | n$^{14}$N | -1/2 + 0 = -1/2 | 1/2 | [68,69] |
| 19. | $^{16}$N ($2^-$,1) | n$^{15}$N | -1/2 - 1/2 = -1 | 1 | [71] |
| 20. | $^{16}$O (0+,0) | $^4$He$^{12}$C | 0 + 0 = 0 | 0 | [26,61,66] |

*Note.* The properties of nuclei are taken from reviews [15,52-55].